\documentclass[10pt,a4paper]{article}
\usepackage{epsfig}
\usepackage{graphicx}
\usepackage{color}

\def\Journal#1#2#3#4{{\it #1} {\bf #2}, #3 (#4)}
\def\book#1#2#3{{\it #1} (#2, #3)}
\def\eprint#1{e-print #1}

\def\PLB{{Phys. Lett.}  B}
\def\PRL{Phys. Rev. Lett.}
\def\PRD{{Phys. Rev.} D}
\def\ZPC{{Z. Phys.} C}
\def\PRP{Phys. Rep.}

\def\JPG{J. Phys. G}
\def\IJMPA{Int. J. Mod. Phys. A}
\def\ZPC{Z. Phys. C}
\def\CPC{Comput. Phys. Commun.}
\def\EPJC{Eur. Phys. J. C}
\def\RMP{Rev. Mod. Phys.}

\def\ibid{\textit{ibid.}}

\begin{document}

\title{Model-independent estimates for the Abelian $Z'$ boson at modern hadron colliders}
\author{Alexey Gulov\footnote{gulov@dsu.dp.ua}~ and
Andrey Kozhushko\footnote{a.kozhushko@yandex.ru}\\
{\it Dnipropetrovsk National University, Dnipropetrovsk, Ukraine}}
\maketitle
\begin{abstract}
The model-independent constraints on the Abelian $Z'$ couplings
from the LEP data are applied to estimate the $Z'$ production in
experiments at the Tevatron and LHC. The $Z'$ total and partial decay
widths are analyzed. The results are compared with model-dependent
predictions and present experimental data from the Tevatron. If we
assume the 1-2$\sigma$ hints from the LEP data to be a signal of
the Abelian $Z'$ boson, then the Tevatron data constrain the $Z'$
mass between 400 GeV and 1.2 TeV.
\end{abstract}

\section{Introduction}

Searching for signals of new physics beyond the standard model
(SM) is an essential part of experiments at modern colliders. New
phenomena could be discovered through deviations of observed
quantities from the predicted SM background. However, observables
in experiments at hadron colliders can be calculated with
significant theoretical uncertainties coming mainly from the
parton distribution functions of initial states and complicated
structure of hadronic final states. In this situation one can only
hope to discover the most prominent signals in the most clear
processes. This is the reason to pay attention to searching for
resonances of new heavy particles decaying into lepton pairs.

A neutral vector boson ($Z'$ boson) is probably the most
perspective intermediate state in scattering processes of quarks
and leptons which could be discovered in the Tevatron and LHC
experiments. At the parton level it appears in the annihilation
channel, its mass is allowed to be of order 1 TeV by current
experimental constraints, and it is a necessary component of
popular grand unification theories and other models with extended
gauge sector (see \cite{leike,Lang08,Rizzo06} for review).

In general, the accurate description of $Z'$ resonance requires to
consider scattering amplitudes with intermediate virtual states.
But if the resonance is a narrow one, then it can be described in
a more simple way by a small number of convenient characteristics
of the production and the decay of the particle. In this approach
it is enough to set the $Z'$ mass and width, the production
cross-section, and the branching ratio into the final state.
Supposing some numbers for the $Z'$ parameters in various
estimates one could and, in principle, would take into account all
the available experimental constraints on $Z'$ from previous
experiments.

Of course, effects of $Z'$ boson can be calculated in details for
each specific model beyond the SM. Such estimates are widely
presented in the literature \cite{Erler:2009jh,aguila,Erler:2010arxiv}. Some
set of popular $E_6$ based models and left-right models is usually
considered in this approach. However, probing the set we can still
miss the actual $Z'$ model. In this regard, it is useful to
complement model-dependent $Z'$ searching by some kind of
model-independent analysis, i.e. the analysis covering a lot of
models. Almost all of the usually considered models belong to the
models with so-called Abelian $Z'$ boson. In Ref.
\cite{Gulov:2000eh,Gulov:1999ry} we found the relations which hold
in any model containing the Abelian $Z'$ boson and satisfying the
following conditions:
\begin{itemize}
\item
only one neutral vector boson exists at energy scale about 1-10
TeV,
\item
the $Z'$ boson can be phenomenologically described by the
effective Lagrangian \cite{leike,Lang08,Rizzo06} at low energies,
\item
the $Z'$ boson and other possible heavy particles are decoupled at
considered energies, and the theory beyond the $Z'$ decoupling
scale is either one- or two-Higgs-doublet standard model (THDM),
\item
the SM gauge group is a subgroup of possible extended gauge group
of the underlying theory. So, the only origin of possible
tree-level $Z'$ interactions to the SM vector bosons is the
$Z$--$Z'$ mixing.
\end{itemize}
These relations cover almost all of the usually considered set of
models (see \cite{GulovSkalozub:2009review,GulovSkalozub:2010ijmpa}
for details). They require the same $Z'$
couplings to the left-handed fermion currents within any SM
doublet and the universal absolute value of the $Z'$ couplings to
the axial-vector currents for all the massive SM fermions. The
relations reduce significantly the number of unknown $Z'$
parameters. This allows to constrain the parameters by existing
experiments as well as to predict the quantities used in the
analysis of the Tevatron and LHC experiments.

Recently we summarized the information about $Z'$ couplings to
leptons and quarks which can be extracted from the LEP experiments
\cite{GulovSkalozub:2009review,GulovSkalozub:2010ijmpa}. The $Z'$
coupling to axial-vector currents was constrained by both LEP I
and LEP II $\mu^+\mu^-,\tau^+\tau^-$ data. In different processes
it shows hints at about 1$\sigma$ confidence level (CL) with the
approximately same maximum-likelihood (ML) value. This value can
be used in estimates of observables in the Tevatron and LHC
experiments. As for the couplings to vector currents, the $Z'$
coupling constant to electron can be constrained by the LEP II
$e^+e^-$ data only. Although the backward scattering shows a
signal at the 2$\sigma$ CL, the ML value is outside of the $95\%$
CL interval calculated by the complete set of bins. In this
situation we refrain from using that ML value in our estimates.
Nevertheless, the vector coupling is constrained at $95\%$ CL. The
upper bound on the electron vector coupling agrees closely with
the corresponding upper bound on the axial-vector coupling. This
fact allows us to suppose the rest of vector couplings to be
constrained by the same value, since no evident signals were
discovered in other scattering processes measured by the LEP
collaborations. It is worth to note that all the conclusions
derived from the LEP data are also valid if one considers the THDM
as the low-energy theory instead of the usual minimal SM.

The main goal of the present paper is to obtain estimates for the
$Z'$ parameters used in searching for the narrow resonance by
applying the LEP constraints on the $Z'$ couplings. Both the
minimal SM and the THDM will be considered as the low-energy
theory.

The paper is organized as follows. Sec. \ref{sec:couplings}
contains a necessary information about $Z'$ interactions at low
energies, the relations between the $Z'$ couplings and the limits
on these couplings obtained from the LEP data. In Sec.
\ref{sec:prod} the $Z'$ production cross-section at hadron
colliders is estimated. The bounds on the total and partial decay
widths are presented in Sec. \ref{sec:width}. In Sec.
\ref{sec:discussion} we discuss the application of our results
comparing them, in particular, with the Tevatron experimental data
and model-dependent predictions for the Tevatron and LHC. The
explicit Lagrangian used for the calculations is given in Appendix
\ref{sec:full_lagr}.

\section{Theoretical and experimental constraints on the $Z'$ couplings}\label{sec:couplings}

In this paper we discuss mainly the $Z'$ couplings to the vector
and axial-vector fermion currents described by the Lagrangian
\begin{eqnarray}\label{ZZplagr}
{\cal L}_{Z\bar{f}f}&=&\frac{1}{2} Z_\mu\bar{f}\gamma^\mu\left[
(v^\mathrm{SM}_{fZ}+\gamma^5 a^\mathrm{SM}_{fZ})\cos\theta_0
%+\right.\nonumber\\&&\quad\left.
+(v_f+\gamma^5 a_f)\sin\theta_0 \right]f, \nonumber\\
{\cal L}_{Z'\bar{f}f}&=&\frac{1}{2} Z'_\mu\bar{f}\gamma^\mu\left[
(v_f+\gamma^5 a_f)\cos\theta_0
%-\right.\nonumber\\&&\quad\left.
-(v^\mathrm{SM}_{fZ}+\gamma^5
a^\mathrm{SM}_{fZ})\sin\theta_0\right]f,
\end{eqnarray}
where $f$ is an arbitrary SM fermion state;
$a_f$ and $v_f$ are the $Z'$ couplings to the
axial-vector and vector fermion currents; $\theta_0$
is the $Z$--$Z'$ mixing angle;
$v^\mathrm{SM}_{fZ}$,
$a^\mathrm{SM}_{fZ}$ are the SM couplings of the $Z$-boson. Such a
parametrization is suggested by a number of natural conditions.
First of all, the $Z'$ interactions of renormalizable types are to
be dominant at low energies $\sim m_W$. The non-renormalizable
interactions generated at high energies due to radiation
corrections are suppressed by the inverse heavy mass
$1/m_{Z^\prime}$ (or by other heavier scales $1/\Lambda_i\ll
1/m_{Z^\prime}$) and therefore at low energies can be neglected.
It is also assumed that the $Z^\prime$ is the only neutral vector
boson with the mass $\sim m_{Z^\prime}$.

It is obvious that the Lagrangian (\ref{ZZplagr}) requires the
$Z^\prime$ boson to enter the theory as a gauge field through
covariant derivatives with a corresponding charge. This idea
allows also to introduce $Z'$ couplings to SM scalar and vector
fields. Although the latter couplings are inessential in the
analysis of the $Z'$ production cross-section in fermion
collisions, they contribute to the $Z'$ width. We assume that the
${\rm SU}(2)_L\times{\rm U}(1)_Y$ gauge group of the SM is a
subgroup of the GUT group. In this case, a product of generators
associated with the SM subgroup is a linear combination of these
generators. As a consequence, all the structure constants
connecting two SM gauge bosons with $Z^\prime$ have to be zero.
Hence, the $Z'$ interactions to the SM gauge fields at the tree
level are possible due to a $Z$--$Z'$ mixing only.

We will consider both the SM and the THDM
as the low-energy theory. The explicit Lagrangian
describing $Z'$ couplings to the SM fields can be found in
Appendix \ref{sec:full_lagr}.

The parameters $a_f$, $v_f$, and $\theta_0$ must be fitted in
experiments. In a particular model, one has some specific values
for them. In case when the model is unknown, these parameters
remain potentially arbitrary numbers. In most investigations they
are usually considered as independent ones. However, this is not
the case if one assumes that the underlying extended model is a
renormalizable one. In Refs. \cite{Gulov:2000eh,Gulov:1999ry} it
was shown that these parameters are correlated as
\begin{equation} \label{grgav}
v_f - a_f= v_{f^*} - a_{f^*}, \qquad a_f = T_{3f}
\tilde{g}\tilde{Y}_\phi,
\end{equation}
where $f$ and $f^*$ are the partners of the $SU(2)_L$ fermion
doublet ($l^* = \nu_l, \nu^* = l, q^*_u = q_d$ and $q^*_d = q_u$),
$T_{3f}$ is the third component of weak isospin, and
$\tilde{g}\tilde{Y}_\phi$ determines the $Z'$ interactions to the
SM scalar fields (see Appendix for details). The parameter
$\tilde{g}\tilde{Y}_\phi$ defines also the $Z$--$Z'$ mixing angle
in (\ref{ZZplagr}).

As it was discussed in
\cite{GulovSkalozub:2009review,GulovSkalozub:2010ijmpa}, the
relations (\ref{grgav}) cover a popular class of models based on
the ${\rm E}_6$ group (the so called LR, $\chi$-$\psi$ models) and
other models, such as the Sequential SM \cite{SeqSM}. Thus, they
describe correlations between $Z'$ couplings for a wide set of
models beyond the SM. That is the reason to call the relations
model-independent ones.

The couplings of the Abelian $Z'$ to the axial-vector fermion
current have a universal absolute value. The value is proportional
to the $Z'$ coupling to scalar fields. Then, the $Z$--$Z'$ mixing
angle $\theta_0$ can be also determined by the axial-vector
coupling.

At low energies the $Z'$ couplings enter the cross-section
together with the inverse $Z'$ mass, so it is convenient to
introduce the dimensionless couplings
\begin{equation}\label{avbar}
\bar{a}_f=\frac{m_Z}{\sqrt{4\pi}m_{Z'}}a_f,\quad
\bar{v}_f=\frac{m_Z}{\sqrt{4\pi}m_{Z'}}v_f,
\end{equation}
which are constrained by experiments. Since the axial-vector
coupling is universal, we will use the notation
\begin{equation}\label{RGrel1}
\bar{a} = \bar{a}_d = \bar{a}_{e^-} = -\bar{a}_u = -\bar{a}_{\nu}.
\end{equation}
Then the $Z$--$Z'$ mixing is
\begin{equation}\label{RGrel2}
\theta_0 \approx -2\bar{a}\frac{\sin \theta_W \cos
\theta_W}{\sqrt{\alpha_{\rm em}}} \frac{m_Z}{m_{Z'}}.
\end{equation}
It also follows from (\ref{grgav}) that for each fermion doublet
only one vector coupling is independent:
\begin{equation}\label{RGrel3}
\bar{v}_{f_d} = \bar{v}_{f_u} + 2 \bar{a}.
\end{equation}
As a result, $Z'$ couplings can be parameterized by seven
independent constants $\bar{a}$, $\bar{v}_u$, $\bar{v}_c$,
$\bar{v}_t$, $\bar{v}_e$, $\bar{v}_\mu$, $\bar{v}_\tau$.

Recently we obtained limits on $Z'$ couplings from the LEP I and
LEP II data
\cite{GulovSkalozub:2009review,GulovSkalozub:2010ijmpa}. We found
some hints of $Z'$ boson at 1-2$\sigma$ CL. Namely, the constants
$\bar{a}$ and $\bar{v}_e$ show non-zero ML values. The
axial-vector coupling $\bar{a}$ can be constrained by the LEP I
data (through the mixing angle) and by the LEP II
$e^+e^-\to\mu^+\mu^-,\tau^+\tau^-$ data. The corresponding ML
values are very close to each other. This value
\begin{equation}\label{MLV}
\bar{a}^2=1.3\times 10^{-5}
\end{equation}
will be used in our estimates. The 95\% CL interval was also
obtained by the experimental data:
\begin{equation}\label{CLI1}
0< \bar{a}^2 < 3.61\times 10^{-4}.
\end{equation}

The electron vector coupling $\bar{v}_e$ can be constrained by the
LEP II $e^+e^-\to e^+ e^-$ data. An evident non-zero
ML value occurred in fits taking into account the
backward scattering bins only. Those fits showed 2$\sigma$ signal
of the $Z'$ boson. On the other hand, that ML
value was excluded at 95\% CL by fits including all the bins. This
instability is the reason to refrain from using the
ML value of $\bar{v}_e$ in our estimates. The
$95\%$ CL interval on $\bar{v}_e$ will be taken into account only:
\begin{equation}\label{CLI2}
4\times 10^{-5}< \bar{v}_e^2 <1.69\times 10^{-4}.
\end{equation}

Other $Z'$ coupling constants cannot be severely constrained by
existing data. Among them $\bar{v}_u$, $\bar{v}_c$, and
$\bar{v}_\mu$ play an important role in the process $q\bar{q}\to
Z'\to\mu^+\mu^-$ which is most perspective to discover the $Z'$
resonance. Taking into account that no evident signals of new
physics were found by the LEP collaborations in the processes
involving quarks, muons and tau-leptons, we constrain the values
of $\bar{v}_u$, $\bar{v}_c$, $\bar{v}_t$, $\bar{v}_\mu$, and
$\bar{v}_\tau$ by the widest interval from the 95\% CL intervals
for $\bar{a}$ and $\bar{v}_e$:
\begin{equation}\label{CLI3}
0< \bar{v}_{\mathrm{other}\ f}^2 <4\times 10^{-4}.
\end{equation}

The knowledge of possible values of the $Z'$ couplings allows to
estimate the $Z'$ production cross-section at the LHC and Tevatron
and the $Z'$ decay width without specifying the model beyond the
SM.

\section{$Z'$ production cross-section}\label{sec:prod}

\begin{figure}
\begin{minipage}[b]{0.33\linewidth}
\psfig{file=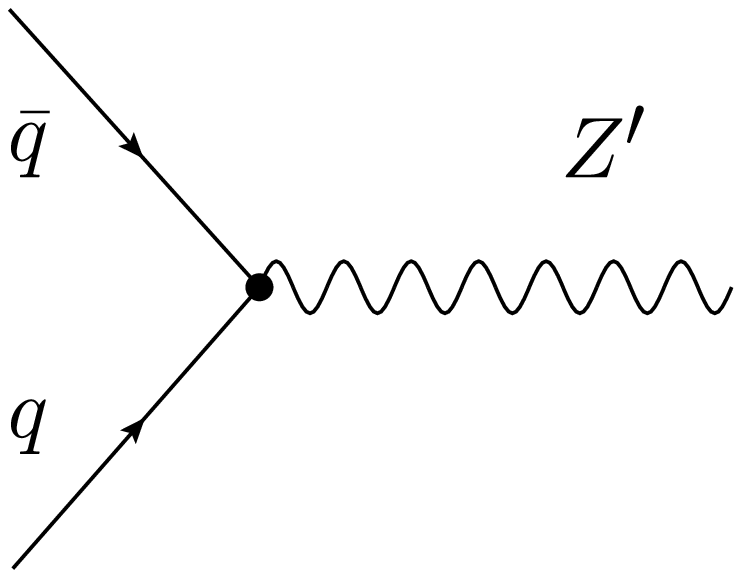,width=40mm}
 \caption{$Z'$ production at the parton level.}
 \label{fig:Prod}
\end{minipage}
\hfil
\begin{minipage}[b]{0.5\linewidth}
\psfig{file=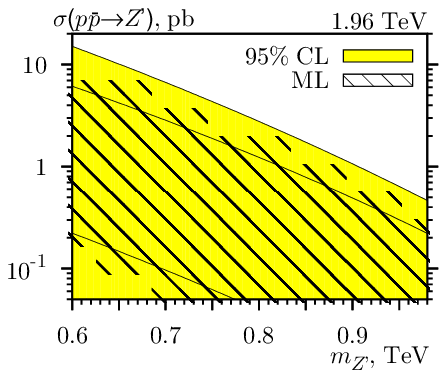,width=60mm}
 \caption{$Z'$ production cross-section vs. $m_{Z'}$ in $p\bar{p}$ collisions
 at $\sqrt{S} = 1.96$ TeV.
 The filled area corresponds to the 95\% CL estimate, and the hatched area is for the
 ML estimate.}
 \label{fig:tevatron_prodcs}
\end{minipage}
\end{figure}

In modern experiments $Z'$ bosons are expected to be produced in
proton-antiproton collisions $p\bar{p} \to Z'$ (Tevatron) or
proton-proton collisions $pp \to Z'$ (LHC). At the parton level
both the processes are described by the annihilation of a
quark-antiquark pair, $q\bar{q} \to Z'$ (Fig. \ref{fig:Prod}). The
$Z'$ production cross-section is the result of integration of the
partonic cross-section $\sigma_{q\bar{q}\to Z'}$ with the parton
distribution functions:
\begin{eqnarray}
\sigma_{AB} &=&
\sum_{q,\bar{q}}\int_{0}^{1}dx_{q}\int_{0}^{1}dx_{\bar{q}} \, f_{q,A}
(x_q,Q^2)f_{\bar{q},B}(x_{\bar{q}},Q^2)
\nonumber\\
&&\times \sigma_{q\bar{q}\to Z'}(m_{Z'}, x_q k_A, x_{\bar{q}} k_B),
\end{eqnarray}
where $A$, $B$ mark the interacting hadrons ($p$ or $\bar{p}$)
with the four-momenta $k_A$, $k_B$; $f_{q,A}$ is the parton
distribution function for the parton $q$ in the hadron $A$ with
the momentum fraction $x_q$ ($0 \leq x_q \leq 1$) at the energy
scale $Q^2$. In our case $Q^2 = m_{Z'}^2$. We use the parton
distribution functions provided by the MSTW PDF package
\cite{mstw}.

The production cross-section is determined by quadratic
combinations of the $Z'$ couplings to quarks,
\begin{eqnarray}
\label{eq:factorcs}
\sigma_{AB} &=& \bar{a}^2 \sigma_{\bar{a}^2} + \bar{a}\bar{v}_u \sigma_{\bar{a}\bar{v}_u} + \bar{v}_u^2 \sigma_{\bar{v}_u^2} + \bar{a}\bar{v}_c \sigma_{\bar{a}\bar{v}_c} \nonumber\\
&& + \bar{v}_c^2 \sigma_{\bar{v}_c^2} + \bar{a}\bar{v}_t \sigma_{\bar{a}\bar{v}_t} + \bar{v}_t^2 \sigma_{\bar{v}_t^2}.
\end{eqnarray}
where relations (\ref{RGrel1})--(\ref{RGrel3}) are taken into
account. The factors $\sigma$ depend on $m_{Z'}$, the process type
(proton-proton or proton-antiproton collision), and the beam
energy. The factors $\sigma_{\bar{a}\bar{v}_c}$,
$\sigma_{\bar{v}_c^2}$, $\sigma_{\bar{a}\bar{v}_t}$ and
$\sigma_{\bar{v}_t^2}$ are small compared to $\sigma_{\bar{a}^2}$,
$\sigma_{\bar{a}\bar{v}_u}$ and $\sigma_{\bar{v}_u^2}$ and their
contributions to the cross-section can be neglected.

We take into account the 90\% CL uncertainties of the parton
distribution functions provided by the MSTW PDF package. Finally,
the production cross-section reads:
\begin{eqnarray}
\label{eq:factorcsshort} \sigma & = & \bar{a}^2 \sigma_{\bar{a}^2}
+ \bar{a}\bar{v}_u \sigma_{\bar{a}\bar{v}_u} +
\bar{v}_u^2 \sigma_{\bar{v}_u^2} \pm \Delta \sigma^{\rm pdf}, \nonumber\\
\Delta \sigma^{\rm pdf} & = & \bar{a}^2 \Delta \sigma^{\rm
pdf}_{\bar{a}^2} + \bar{a}\bar{v}_u \Delta \sigma^{\rm
pdf}_{\bar{a}\bar{v}_u} + \bar{v}_u^2 \Delta \sigma^{\rm
pdf}_{\bar{v}_u^2}.
\end{eqnarray}

Due to the existence of the ML value for the
axial-vector coupling we perform two different estimates for the
production cross-section:
\begin{itemize}
\item \textit{95\% CL estimate}. In this scheme both the couplings $\bar{a}$
and $\bar{v}_u$ are varied in their 95\% CL intervals
(\ref{CLI1}), (\ref{CLI3}). Then the production cross-section lies
inside of the interval between zero and some maximal value. The
maximal value is reached when both the couplings $\bar{a}$ and
$\bar{v}_u$ are of the same sign and take their maximal values:
$\bar{a}=\sqrt{3.61} \times 10^{-2}$, $\bar{v}_u = 0.02$. The
uncertainty from the parton distribution functions should be also
added. This estimate leads to the widest interval of possible
values of the production cross-section.

\item \textit{Maximum-likelihood estimate}.
In this approach the axial-vector coupling is substituted
by its ML value $\bar{a}=\sqrt{1.3 \times
10^{-5}}$. The vector coupling $\bar{v}_u$ is varied in its 95\%
CL interval. If one chooses the positive value of the axial-vector
coupling, then the minimal value of the cross-section corresponds
to $\bar{v}_u\simeq -0.02$ whereas the maximal value is reached at
$\bar{v}_u\simeq 0.02$. The obtained interval should be also
enlarged by $\Delta \sigma^{\rm pdf}$. This estimate gives a more
narrow interval for the production cross-section which can be
considered as an `optimistic' scenario to discover the $Z'$ boson.
\end{itemize}

\begin{figure}[!h]
\centering{\psfig{file=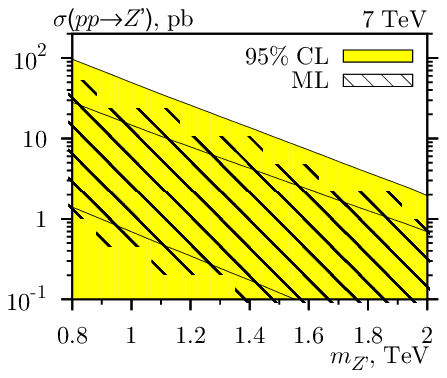,width=60mm}
\psfig{file=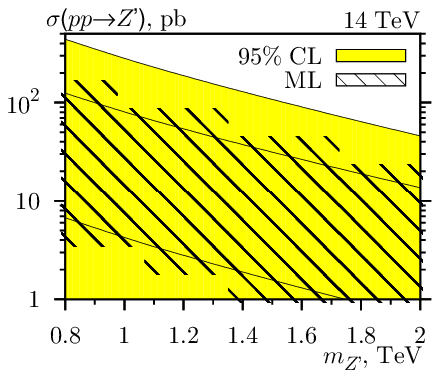,width=60mm}}
\caption{$Z'$ production cross-section vs. $m_{Z'}$ in $pp$ collisions
 at $\sqrt{S} = 7$ TeV and $\sqrt{S} = 14$ TeV.
 The filled area corresponds to the 95\% CL estimate, and the hatched area is for the
 ML estimate.}
\label{fig:lhc_prodcs}
\end{figure}

The estimates for the $Z'$ production cross-section in
proton-antiproton collisions at the Tevatron and in proton-proton
collisions at the LHC are shown in Figs. \ref{fig:tevatron_prodcs}
and \ref{fig:lhc_prodcs}, respectively. In the LHC case the
$\sqrt{S}$ value is taken to be 7 TeV and 14 TeV, corresponding to
the current and expected energies. The $Z'$ mass is chosen to be
from 600 to 980 GeV for the Tevatron process and from 800 to 2000
GeV for the LHC processes. At these masses it is possible to
perform direct searches, and the boson production rate is not
suppressed by the parton density effects.

\section{$Z'$ width}\label{sec:width}

\begin{figure}
\centering{\psfig{file=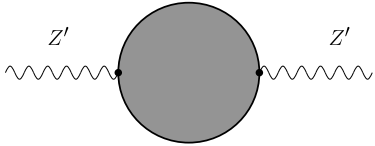,width=50mm}}
\caption{One-particle-irreducible correction to $Z' \to Z'$.} \label{fig:Width}
\end{figure}

The $Z'$ decay width $\Gamma_{Z'}$ can be calculated by using the
optical theorem:
\begin{eqnarray}
\Gamma_{Z'} = -\frac{{\rm Im}\,G(m_{Z'}^2)}{m_{Z'}},
\end{eqnarray}
where $G(p^2)$ is the two-point one-particle-irreducible Green's
function corresponding to the diagram in Fig. \ref{fig:Width}. We
compute $\Gamma_{Z'}$ at the one-loop level with the help of the
FeynArts, FormCalc and LoopTools software \cite{FeynArts,FormCalc}.
The Feynman diagrams with
internal $Z'$ lines as well as the Passarino-Veltman integrals of
type $A$ give no contribution to the result, since they are real.
The rest of diagrams correspond to different channels of $Z'$
decay. As a result, we obtain also all the partial widths
corresponding to $Z'$ decays into two SM particles.

All the $Z'$ couplings to the SM scalar and vector bosons can be
determined by the universal axial-vector constant $a_f$ and can be
constrained. Then the partial widths corresponding to $Z'$ decays
into scalar and vector bosons are proportional to $a_f^2$. As for
the fermionic decays, the width can be written in the form
\begin{equation}\label{eq:factorwidth}
\Gamma_{Z'\to \bar{f}f} = a_f^2 \Gamma_{{a}_f^2} + {a}_f {v}_f
\Gamma_{{a}_f {v}_f} + {v}_f^2 \Gamma_{{v}_f^2}.
\end{equation}
The factors $\Gamma_{{a}_f^2}$, $\Gamma_{{a}_f {v}_f}$ and
$\Gamma_{{v}_f^2}$ are proportional to $m_{Z'}$. Expressing eq.
(\ref{eq:factorwidth}) through the constants (\ref{avbar}) one can
see that the width is proportional to $m_{Z'}^3$ and quadratic
combinations of couplings $\bar{a}$, $\bar{v}_f$. Thus it is
convenient to introduce quantity
\begin{equation}
\label{eq:gammatilde} \tilde{\Gamma} = \Gamma_{Z'} \times \left(
\frac{1 \textrm{ TeV}}{m_{Z'}} \right)^3,
\end{equation}
which is independent of $m_{Z'}$ in our estimates.

To calculate $\tilde{\Gamma}$ numerically one has to choose values
of the unknown masses of the SM scalar particles. If the minimal
SM is considered as the low-energy theory, the only unknown mass
is the Higgs boson mass $m_h$. The modern constraints on its value
indicate that it is quite heavy, $m_h \geq 114$ GeV. The
contribution to the decay width from the scalar sector appears to
be two or three orders of magnitude lower than the leading
contribution from the fermionic decay channel. So the decay widths
calculated at different values of $m_h$ are practically
indistinguishable. In this regard, we present the results obtained
for $m_h=125$ GeV.

When the THDM is considered, the scalar sector has six free
parameters that can be expressed in terms of the masses $m_h$,
$m_H$, $m_{A_0}$, $m_{H^\pm}$ and the mixing angles $\tan \alpha$,
$\tan \beta$ (see Appendix \ref{sec:full_lagr} for details).
Because of the large number of physical scalar fields the
estimates for the $Z'$ width within the THDM can deviate from the
results obtained in the case of the minimal SM. In order to obtain
the most significant difference, we choose $H^{\pm}$ and $A_0$ to
be as light as it is allowed by the LEP constraints \cite{RPP},
namely
\begin{equation}
m_{H^\pm} = 81 \textrm{ GeV}, \quad m_{A_0} = 92 \textrm{ GeV}.
\end{equation}
The $h$ and $H$ masses are set to
\begin{equation}
m_h = m_H = 125 \textrm{ GeV}
\end{equation}
just like in the SM case. The dependence of $\tilde{\Gamma}$ on
the mixing angles is negligibly weak. We take $\tan \beta = 2$,
which respects the LEP constraints. The $\tan \alpha$ value is set
to $0.75$.

The decay width is estimated in two schemes which are similar to
the case of the production cross-section:
\begin{itemize}
\item \textit{95\% CL estimate}. In this scheme the coupling constants $\bar{a}$
and $\bar{v}_f$ are varied in their 95\% CL intervals
(\ref{CLI1}), (\ref{CLI3}). The minimal value of the width is
calculated at $\bar{a}=\bar{v}_u=\bar{v}_{\mu,\tau}=0$,
$\bar{v}_e=\pm\sqrt{0.4}\times 10^{-2}$. The maximal value
is realized when all the
couplings are at their maximal absolute values, $\bar{a}$ and
$\bar{v}_{u,c,t}$ are of the same sign, while
$\bar{v}_{e,\mu,\tau}$ have the opposite sign with respect to
$\bar{a}$: $\bar{a}=\pm \sqrt{3.61}\times 10^{-2}$,
$\bar{v}_{u,c,t}=\pm 0.02$, $\bar{v}_{\mu,\tau}=\mp 0.02$,
$\bar{v}_e=\mp \sqrt{1.69}\times 10^{-2}$.

\item \textit{Maximum-likelihood estimate}.
We set $\bar{a}=\sqrt{0.13}\times 10^{-2}$ and vary $v_{f}$ in
their 95\% CL intervals. We choose the positive value of
$\bar{a}$, so the minimum value of the width corresponds to
$\bar{v}_e = \sqrt{0.4}\times 10^{-2}$ and $\bar{v}_f = -\bar{a}_f
\tilde{\Gamma}_{\bar{a}_f
\bar{v}_f}/2\tilde{\Gamma}_{\bar{v}_f^2}$ ($f = \mu, \tau,u,c,t$).
The maximum value is reached at $\bar{v}_{u,c,t}=0.02$,
$\bar{v}_{\mu,\tau}=-0.02$, $\bar{v}_e=-\sqrt{1.69}\times
10^{-2}$.
\end{itemize}

\begin{figure}[!h]
\centering{\psfig{file=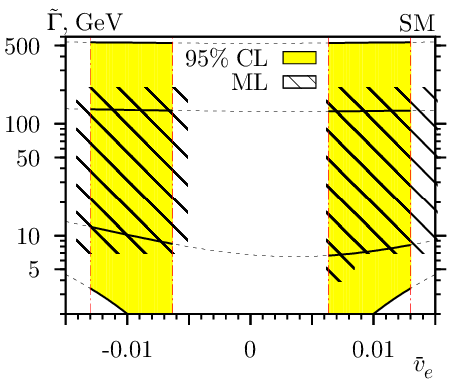,width=60mm}
\psfig{file=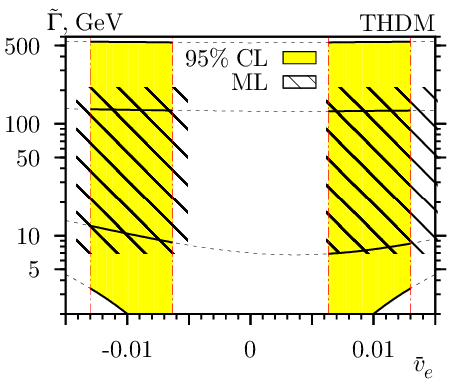,width=60mm}} \caption{The $Z'$ width
(\ref{eq:gammatilde}) versus $\bar{v}_e$ for the SM and THDM
cases. The filled areas represent the 95\% CL estimate, whereas
the hatched areas represent the ML estimate. The inner vertical
dot-dashed lines stand for the minimum 95\% CL value of
$\bar{v}_e$ from the special one-parameter fit of the LEP II data,
the outer ones depict the maximum 95\% CL value of $\bar{v}_e$
from the general two-parameter fit of the LEP II data.}
\label{fig:gammavsve}
\end{figure}

\begin{figure}[!h]
\centering{\psfig{file=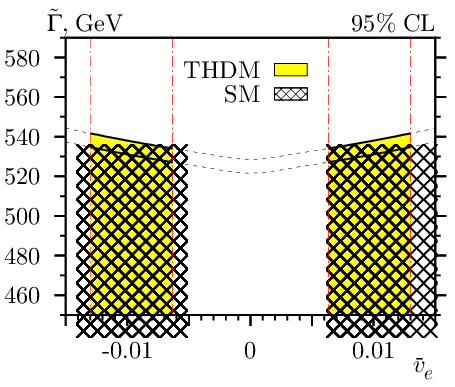,width=60mm}
\psfig{file=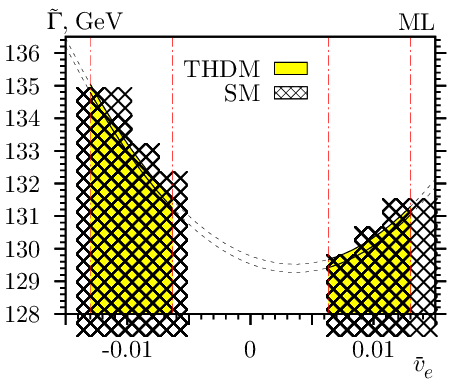,width=60mm}} \caption{The
$\tilde{\Gamma}$ estimates versus $\bar{v}_e$ for the SM and THDM
cases. The filled areas represent the estimate for the THDM case,
and the hatched areas represent the estimate for the SM case. The
meaning of the vertical dot-dashed lines is the same as in Fig.
\ref{fig:gammavsve}.} \label{fig:gammasmthdm}
\end{figure}

The $Z'$ width (\ref{eq:gammatilde}) is plotted in Fig.
\ref{fig:gammavsve} as the function of $\bar{v}_e$. The minimal SM
and the THDM lead to slightly different bounds depicted in Fig.
\ref{fig:gammasmthdm}.

Since we chose the positive ML value of the
axial-vector coupling $\bar{a}$, we obtain asymmetric domain in
the parameter space within the ML estimate as it
is seen in Figs. \ref{fig:gammavsve} and \ref{fig:gammasmthdm}.
This asymmetry arises from the term $\bar{a}\bar{v}_e
\Gamma_{\bar{a}\bar{v}_e}$ in (\ref{eq:factorwidth}). Of course,
the sign of $\bar{a}$ is not constrained by the experimental data,
so the sign of the vector coupling should be considered as the
relative sign with respect to the axial-vector coupling. For the
electron vector coupling the $2\sigma$ hint was observed
\cite{GulovSkalozub:2010ijmpa}. This allows to exclude the area
near $\bar{v}_e = 0$ shown in the figures.

Consider an example of usage of the obtained estimates. Let us
assume that the $Z'$ mass is of order $1$--$2$ TeV, for instance
$m_{Z'} = 1.5$ TeV, so $Z'$ production rate in the LHC and
Tevatron processes is non-negligible and the direct searches are
possible. The ML value $\tilde{\Gamma}\approx 50$
GeV leads to the total decay width $\Gamma_{Z'}=169$ GeV. Thus we
can expect the $Z'$ resonance compatible with the narrow width
approximation (NWA), $\Gamma_{Z'}^2/m_{Z'}^2=0.013\ll 1$. However,
one has to keep in mind that $\Gamma_{Z'}\approx m_{Z'}$ is not
excluded at the 95\% CL. The extremely narrow resonances with
$\Gamma_{Z'}\approx 1$ GeV are also not excluded.

It is also useful to estimate the partial decay widths of the $Z'$
boson. In this analysis we take the ML value of
the axial-vector coupling $\bar{a}=\sqrt{0.13}\times 10^{-2}$ and
vary other couplings in their 95\% CL intervals. The results are
presented as the plots in which a partial width is depicted versus
the total width. In this way the branching ratios can be easily
obtained.

\begin{figure}[!h]
\begin{minipage}{0.5\linewidth}
\centering{\psfig{file=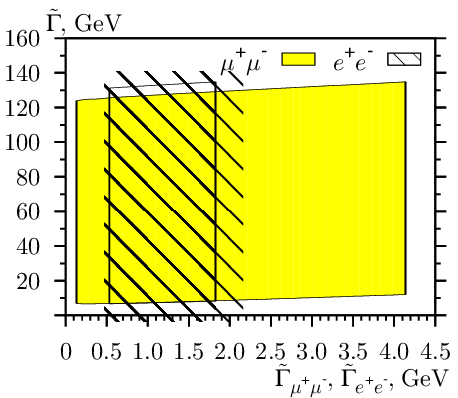,width=60mm}\\(a)}
\end{minipage}
\begin{minipage}{0.5\linewidth}
\centering{\psfig{file=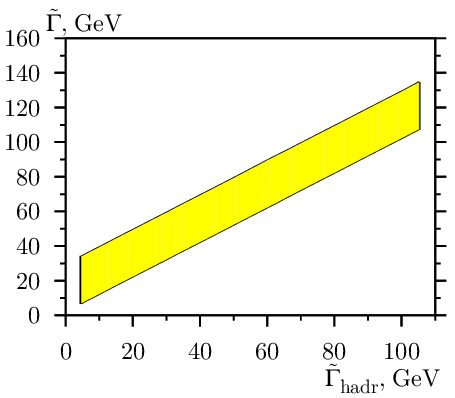,width=60mm}\\(b)}
\end{minipage}
\caption{(a) The ML estimates on the $\tilde{\Gamma}$ versus
$\tilde{\Gamma}_{e^{+}e^{-}}$ and $\tilde{\Gamma}_{\mu^+ \mu^-}$
plane. The filled area is for dimuon channel, and the hatched area
represents the dielectron channel. (b) The ML domain on the
$\tilde{\Gamma}$--$\tilde{\Gamma}_{\rm hadr}$ plane.}
\label{fig:totalvseemumuhadr}
\end{figure}
The partial decay widths for the electron-positron, muon-antimuon,
and quark-antiquark channels are shown in Fig.
\ref{fig:totalvseemumuhadr}. On these plots, the difference
between the SM and the THDM case is negligible. As it is seen, the
branching ratio for the electron-positron decay channel can be
expected in the wide interval
\begin{eqnarray}
\label{eq:br_ee} 0.004 \leq BR(Z' \to e^+ e^-) \leq 0.21.
\end{eqnarray}
Here, the minimal value corresponds to $\bar{v}_e = 0$, whereas
the maximal value is reached at $\bar{v}_e = -\sqrt{1.69} \times
10^{-2}$. The significant difference between the estimates for
$\tilde{\Gamma}_{e^{+}e^{-}}$ and
$\tilde{\Gamma}_{\mu^{+}\mu^{-}}$ is caused by the fact that the
$Z'$ vector coupling to electron is much better constrained by the
LEP II data that the muon one. The decay into quark-antiquark
pairs can be the dominant decay channel. The corresponding
probability can amount to 98\%.

Considering the $Z'$ partial widths, one can find a significant
distinction between the SM and THDM in the scalar sector. Since
$\bar{a}$ is the only $Z'$ coupling entering the scalar and vector
contributions to $\Gamma_{Z'}$, there is the ML
value of the partial decay width into two SM bosons (vectors or
scalars). In the SM case, $\tilde{\Gamma}_{\rm bosons}=0.27$ GeV.
In the THDM case, $\tilde{\Gamma}_{\rm bosons}=0.53$ GeV. The
corresponding branching ratios are less than 2.5\%.

\section{Discussion}
\label{sec:discussion}

The recent experiments at the LEP gave some hints of the Abelian
$Z'$ boson. Although these hints correspond to 68-95\% CL, they
can be used as a beacon showing the most optimistic scenario to
find $Z'$ boson with a mass near 1 TeV. It is interesting to
speculate about the question how can those hints look like at
Tevatron and LHC experiments. Taking the LEP ML value of the
axial-vector coupling we can give predictions under the assumption
that a signal of the Abelian $Z'$ boson has been probably observed
in the LEP data. This estimate, called ML scheme, represents the
most bold expectations concerning the Abelian $Z'$ boson. Of
course, such predictions do not exclude $Z'$ boson with weaker
axial-vector couplings.

On the other hand the 95\% CL bounds on possible $Z'$ couplings to
the SM particles are left behind the LEP experiments. Taking these
bounds for all the $Z'$ couplings we can exclude some values of
the observables at hadron colliders. In this scheme the values
outside of the predicted intervals are forbidden for the Abelian
$Z'$ boson. Being measured in experiments, such values have to be
interpreted as a signal of new physics which is something else
than the $Z'$ boson. For example, considering the $Z'$ width, we
can expect $\Gamma_{Z'} \times (1\textrm{ TeV}/m_{Z'})^3 \simeq
10-150$ GeV from the ML estimate, and we can think
about the NWA for $m_{Z'}\le 2$ TeV. On
the other hand, only extremely narrow resonances, $\Gamma_{Z'}
\times (1\textrm{ TeV}/m_{Z'})^3<1$ GeV, and extremely wide
resonances, $\Gamma_{Z'} \times (1\textrm{ TeV}/m_{Z'})^3>500$
GeV, can be surely excluded at the 95\% CL. Thus, waiting for a
narrow $Z'$ resonance at hadron colliders we have to keep in mind
that a more rich $Z'$ phenomenology is still allowed by existing
data.

\begin{figure}[!h]
\begin{minipage}{0.5\linewidth}
\centering{\psfig{file=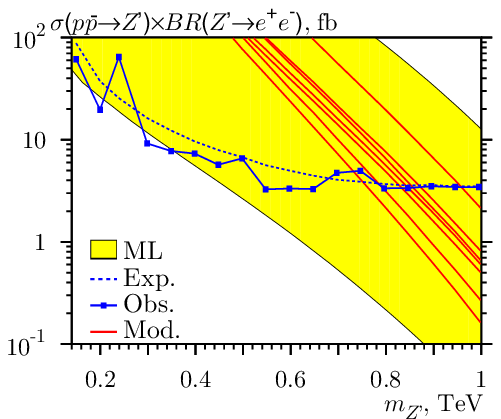,width=60mm}\\(CDF,
dielectron channel)}
\end{minipage}
\begin{minipage}{0.5\linewidth}
\centering{\psfig{file=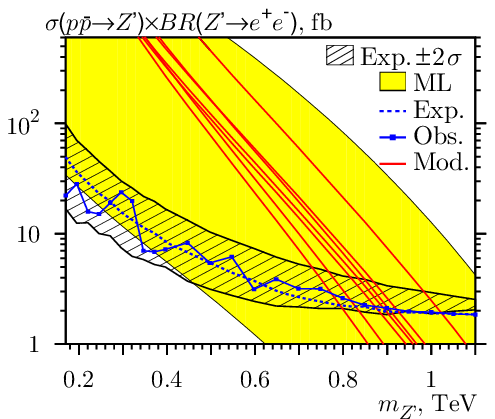,width=60mm}\\(D0, dielectron
channel)}
\end{minipage}
\begin{minipage}{0.5\linewidth}
\centering{\psfig{file=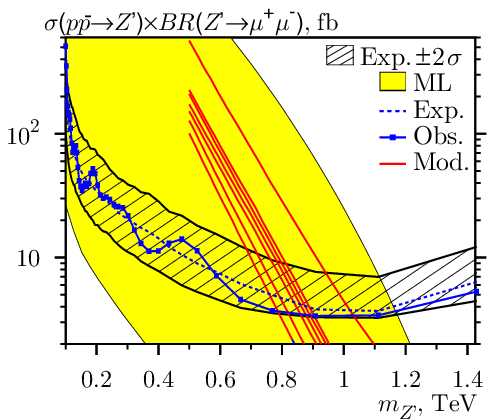,width=60mm}\\(CDF, dimuon
channel)}
\end{minipage}
\caption{The comparison between the Tevatron results and the ML
estimates of the $Z'$ production in the Drell-Yan process at
$\sqrt{S}=1.96 \textrm{ TeV}$. In all plots the filled areas
represent the ML estimates. The experimentally obtained upper
limits on the $Z'$ contribution are taken from
\cite{CDF_zpr,CDF_zpr_mumu,D0_zpr}: the expected and observed 95\%
CL upper limits are depicted by the dashed lines and line charts,
respectively, and the hatched areas are the 2$\sigma$ standard
deviation bands for the expected values. The predictions from the
popular models \cite{CDF_zpr,CDF_zpr_mumu,D0_zpr} are plotted as
solid red lines, the corresponding models are $Z_{\rm I}'$,
$Z_{\rm sec}'$, $Z_{\rm N}'$, $Z_{\psi}'$, $Z_{\chi}'$,
$Z_{\eta}'$ and SSM $Z'$ from the left to the right. }
\label{fig:sbr_tevatron}
\end{figure}

%Tevatron
Now let us present the ML estimate for the
Drell-Yan cross-section for the Tevatron and LHC experiments. As
it was mentioned, in this case the NWA can be applied and the $Z'$
contribution to the cross-section of the $pp \, (p\bar{p}) \to
l\bar{l}$ process reads $\sigma(pp \, (p\bar{p}) \to Z') \times
BR(Z' \to l\bar{l})$ where the branching ratio can be extracted
from the total and partial $Z'$ decay widths. The experimental
bounds on the $Z'$ contribution to the Drell-Yan process at the
Tevatron are available in \cite{CDF_zpr,CDF_zpr_mumu,D0_zpr}
together with the predictions from popular $Z'$ models. The
comparison between those results and our ML
estimate for $\sigma(p\bar{p} \to Z' \to e^+ e^-,\, \mu^+ \mu^-)$
is presented in Figs. \ref{fig:sbr_tevatron}. We can conclude from
both the D0 and CDF limits that the $Z'$ hints from the LEP data
can be the Abelian $Z'$ boson  with the mass between 400 GeV and
1.2 TeV. Our model-independent results cover all the popular $Z'$
models. We can also conclude that the model-independent lower
bound on the $Z'$ mass is still about 400 GeV whereas the popular
models give the lower bound of order $800-900$ GeV.

\begin{figure}[t]
\centering{\psfig{file=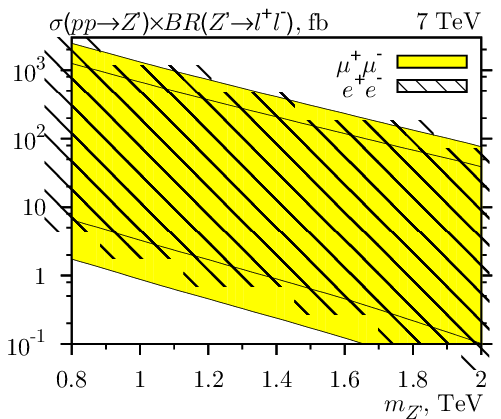,width=60mm}
\psfig{file=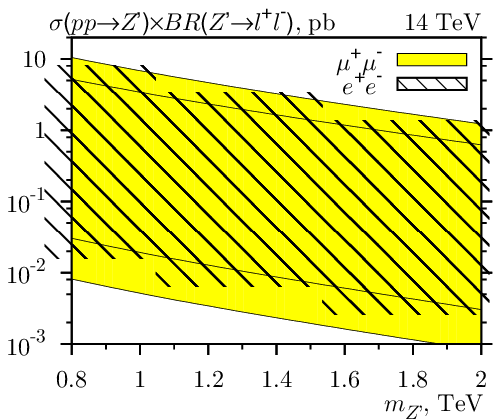,width=60mm}}
\caption{The ML domain for $\sigma(pp \to Z') \times BR(Z' \to e^+ e^-)$
(hatched area) and
$\sigma(pp \to Z') \times BR(Z' \to \mu^+ \mu^-)$ (filled area)
at $\sqrt{S}=7 \textrm{ TeV}$ and $14 \textrm{ TeV}$.}
\label{fig:sbr_lhc}
\end{figure}

%(LHC)
It is straightforward to carry out similar calculations for $pp
\to Z' \to l^+ l^-$ processes at the LHC. The ML
domains are presented in Figs. \ref{fig:sbr_lhc}. The cross-section
values are plotted for the $Z'$ mass up to 2 TeV. For
higher masses the validity of the NWA is not guaranteed even for
the ML estimate. Let us compare the results to the
ones presented in \cite{Dittmar2003}. In Fig. 3 of Ref.
\cite{Dittmar2003} the number of $pp \to Z' \to l^+ l^-$ events
for 100 fb$^{-1}$ of integrated luminosity at $\sqrt{S}=14
\textrm{ TeV}$ versus $m_{Z'}$ is plotted. The ML
number of $pp \to Z' \to e^+ e^-$ or $\mu^+ \mu^-$ events for this
luminosity can be obtained by multiplying the cross-section values
in the left plot in Fig. \ref{fig:sbr_lhc} by $10^5$. It can be
seen that all the model-dependent predictions from Ref.
\cite{Dittmar2003} are covered by the $e^+ e^-$ ML
domain.

In Table 2 of Ref. \cite{Dittmar2003} the model-dependent
estimates for $\sigma(pp \to Z' \to l^+ l^-) \times \Gamma_{Z'}$
are presented. $m_{Z'}$ is set to 1.5 TeV. The ML
estimate for this observable is easy to calculate using Figs.
\ref{fig:lhc_prodcs} and \ref{fig:totalvseemumuhadr} (a) as
$\sigma(pp \to Z') \times \tilde{\Gamma}^{l^+l^-} \times (m_{Z'}/1
\ \textrm{ TeV})^3$. We obtain $94 \pm 92 \textrm{ pb}\cdot
\textrm{GeV}$ and $210.7 \pm 210.1 \textrm{ pb}\cdot \textrm{GeV}$
for $e^+ e^-$ and $\mu^+ \mu^-$ decay channels, respectively. One
can see that the predictions for the $Z_{\psi}'$ and $Z_{\eta}'$
models ($487 \pm 5 \textrm{ fb}\cdot \textrm{GeV}$ and $630 \pm 20
\textrm{ fb}\cdot \textrm{GeV}$) lie outside the
ML interval for the dielectron channel case, and
the $Z_{\psi}'$ prediction is not covered by the dimuon channel
estimate. This is because $m_{Z'}=1.5$ TeV appears to be quite
heavy to provide exact value of the axial-vector coupling from the
LEP data as it is assumed in the ML scheme. Of
course, the model-dependent results are covered by the 95\% CL
intervals and cannot be excluded by the LEP data.

The model-independent relations for the $Z'$ couplings give a good
possibility to reduce the number of unknown $Z'$ parameters. As a
consequence, the $Z'$ width and the production cross-sections of
the processes at modern hadron colliders can be estimated using
the constraints on the $Z'$ couplings obtained from previous
experiments at LEP. A combined analysis of the LEP, Tevatron and
LHC data seems to be possible.

Our new model-independent results are complementary to the usual
model-dependent schemes. The predictions of all the popular $Z'$
models agree with our model-independent bounds.

Finally the $Z'$ hints observed in the LEP data can be still
hidden as the resonance in the Tevatron experiments. We can expect
this $Z'$ boson with the mass between 400 GeV and 1.2 TeV.

\appendix
\section{Lagrangian}\label{sec:full_lagr}

In this section we adduce the scalar, fermion, Yukawa and gauge
sectors of the Lagrangian that is used for the calculations.

Let $\phi_i$ ($i=1,2$) be two complex scalar doublets:
\begin{equation}
\phi_i^{\mathrm{T}}=\left\{a^+_i, \frac{v_i +b_i +i c_i}{\sqrt{2}}\right\},
\end{equation}
where $v_i$ marks corresponding vacuum expectation values, $a^+_i$
are complex fields, and $b_i$, $c_i$ are real fields. By
diagonalizing the quadratic terms of the scalar potential
$V(\phi_1,\phi_2)$ one obtains the mass eigenstates: two neutral
$CP$-even scalar particles, $H$ and $h$, the neutral $CP$-odd
scalar particle, $A_0$, the Goldstone boson partner of the $Z$
boson, $\chi_3$, the charged Higgs field, $H^\pm$, and the
Goldstone field associated with the $W^\pm$ boson, $\chi^\pm$:
\begin{eqnarray}\label{scalars:rule}
 a^+_1 =& \chi^+\cos\beta -H^+\sin\beta,\quad&
 a^+_2 = H^+\cos\beta +\chi^+\sin\beta, \nonumber\\
 c_1 =& \chi_3\cos\beta -A_0\sin\beta,\quad&
 c_2 = A_0 \cos\beta +\chi_3 \sin\beta, \nonumber\\
 b_1 =& H\cos\alpha -h\sin\alpha,\quad&
 b_2 = h\cos\alpha +H\sin\alpha,
\end{eqnarray}
where
\begin{equation}
\tan\beta =\frac{v_2}{v_1},
\end{equation}
and the angle $\alpha$ is determined by the explicit form of the
potential $V(\phi_1,\phi_2)$. For instance, the $CP$-conserving
potential, which has only $CP$-invariant minima, can be used
\cite{hhguide,santos}:
\begin{eqnarray}\label{L:potential}
V&=& \sum\limits_{i=1}^2\left[
  -\mu^2_i \phi^\dagger_i\phi_i
  +\lambda_i(\phi^\dagger_i \phi_i)^2\right]
 +\lambda_3 (\mbox{Re}[\phi^\dagger_1 \phi_2])^2
 \nonumber\\&&
 +\lambda_4 (\mbox{Im}[\phi^\dagger_1 \phi_2])^2
 +\lambda_5 (\phi^\dagger_1 \phi_1) (\phi^\dagger_2 \phi_2).
\end{eqnarray}
It is consistent with the absence of the tree-level
flavor-changing neutral currents (FCNC's) in the fermion sector.
The corresponding value of $\alpha$ is \cite{santos}
\begin{equation}
\tan 2\alpha = -\frac{v_1 v_2 \left(\lambda_3+\lambda_5\right)}
{\lambda_2 v^2_2-\lambda_1 v^2_1}.
\end{equation}

The $Z'$ coupling to the scalar doublets can be parametrized in a
model independent way as follows \cite{cvetic}:
\begin{eqnarray}\label{L:scalar}
 {\cal L}_\phi
 &=&\sum\limits_{i=1}^2
 \left|\left(\partial_\mu
 -\frac{ig}{2}\sigma_a W^a_\mu
 -\frac{i{g^\prime}}{2}Y_{\phi_i} B_\mu
\right.\right.\left.\left.
 - \frac{i\tilde{g}}{2}\tilde{Y}_{\phi_i} \tilde{B}_\mu
 \right)\phi_i\right|^2,
\end{eqnarray}
where $g$, $g^\prime$, $\tilde{g}$ are the charges associated with
the ${\rm SU}(2)_L$, ${\rm U}(1)_Y$, and the $Z^\prime$ gauge
groups, respectively, $\sigma_a$ are the Pauli matrices,
\begin{eqnarray}
\tilde{Y}_{\phi_i}&=& \left(\begin{array}{cc}
 \tilde{Y}_{\phi_i,1} & 0 \\ 0 & \tilde{Y}_{\phi_i,2}
 \end{array}\right)
\end{eqnarray}
is the generator corresponding to the gauge group of the
$Z^\prime$ boson, and $Y_{\phi_i}$ is the ${\rm U}(1)_Y$
hypercharge. The condition $Y_{\phi_i} =1$ guarantees that the
vacuum is invariant with respect to the gauge group of photon.

The vector bosons, $A$, $Z$, and $Z^\prime$, are related with the
symmetry eigenstates as follows:
\begin{eqnarray}\label{vectors:rule}
 B&\to& A\cos{\theta_W}-(Z\cos{\theta_0}
  -Z^\prime\sin{\theta_0})\sin{\theta_W},
  \nonumber\\
 W_3&\to& A\sin{\theta_W} +(Z\cos{\theta_0}
  -Z^\prime\sin{\theta_0})\cos{\theta_W},
  \nonumber\\
 \tilde{B}&\to& Z\sin{\theta_0}
 +Z^\prime\cos{\theta_0},
\end{eqnarray}
where $\tan\theta_W=g^\prime/g$ is the adopted in the SM value of
the Weinberg angle, and
\begin{eqnarray}\label{ZZpmixing}
 \tan{\theta_0}&=&
 \frac{\tilde{g} m^2_W
 \left(\tilde{Y}_{\phi_1,2}\cos^2\beta
 +\tilde{Y}_{\phi_2,2}\sin^2\beta\right)}
 {g\cos{\theta_W}
 \left(m^2_{Z^\prime}-m^2_W/\cos^2{\theta_W}\right)}.
\end{eqnarray}
As is seen, the mixing angle $\theta_0$ is of order $\sim
m^2_W/m^2_{Z^\prime}$. That results in the corrections of order
$\sim m^2_W/m^2_{Z^\prime}$ to the interactions between the SM
particles. To avoid the tree-level mixing of the $Z$ boson and the
physical scalar field $A_0$ one has to impose the condition
$\tilde{Y}_{\phi_1,2}=
\tilde{Y}_{\phi_2,2}\equiv\tilde{Y}_{\phi,2}$.

The effective low-energy Lagrangian of the fermion-vector
interactions reads \cite{sirlin,caso,cvetic}:
\begin{eqnarray}\label{L:fermion}
 {\cal L}_f
&=&i\sum\limits_{f_L}\bar{f}_L{\gamma^\mu}
 \Big(\partial_\mu
 -\frac{ig}{2}{\sigma_a}W^a_\mu
 -\frac{i g^\prime}{2}{B_\mu}Y_{f_L}
 -\frac{i\tilde{g}}{2}\tilde{B}_\mu\tilde{Y}_{f_L}\Big)f_L
 \nonumber\\
 &&+i\sum\limits_{f_R}\bar{f}_R{\gamma^\mu}
 \Big(\partial_\mu -i g^\prime B_\mu Q_f
 -\frac{i\tilde{g}}{2}\tilde{B}_\mu\tilde{Y}_{R,f}\Big)f_R,
\end{eqnarray}
where the renormalizable type interactions are admitted and the
summation over the all SM left-handed fermion doublets, $f_L
=\{(f_u)_L, (f_d)_L\}$, and the right-handed singlets, $f_R =
(f_u)_R, (f_d)_R$, is understood. $Q_f$ denotes the charge of $f$
in the positron charge units,
\begin{eqnarray}
 \tilde{Y}_{f_L}&=&
 \left(\begin{array}{cc}
 \tilde{Y}_{L,f_u} & 0 \\ 0 & \tilde{Y}_{L,f_d} \end{array}\right),
\end{eqnarray}
and $Y_{f_L}$ equals to $-1$ for leptons and $1/3$ for quarks.

In the present paper we use the $Z'$ couplings to the vector and
axial-vector fermion currents defined as
\begin{equation} \label{eq:av} v_f =
\tilde{g}\frac{\tilde{Y}_{L,f} + \tilde{Y}_{R,f}}{2}, \qquad a_f =
\tilde{g}\frac{\tilde{Y}_{R,f} - \tilde{Y}_{L,f}}{2}.
\end{equation}
The Lagrangian (\ref{L:fermion}) leads to the interactions between
the fermions and the $Z$ and $Z'$ mass eigenstates described by
(\ref{ZZplagr}).

Renormalizable interactions of fermions and scalars are described
by the Yukawa Lagrangian. To avoid the existence of the tree-level
FCNC's one has to ensure that at the diagonalization of the
fermion mass matrix the diagonalization of the scalar-fermion
couplings is automatically fulfilled. In this case the Yukawa
Lagrangian, which respects the ${\rm SU}(2)_L\times{\rm U}(1)_Y$
gauge group, can be written in the form:
\begin{eqnarray}\label{L:Yuk}
{\cal L}_{\rm Yuk} &=&
 -\sqrt{2}\sum\limits_{f_L}\sum\limits_{i=1}^{2}\left\{
 G_{f_d,i}\left[
  \bar{f}_L\phi_i(f_d)_R +(\bar{f}_d)_R \phi^\dagger_i f_L
 \right]
 \right.\nonumber\\&&\left.
+G_{f_u,i}\left[
  \bar{f}_L\phi^c_i(f_u)_R +(\bar{f}_u)_R \phi^{c\dagger}_i f_L
 \right] \right\},
\end{eqnarray}
where $\phi^c_i=i\sigma_2\phi^\ast_i$ is the charge conjugated
scalar doublet, and the Cabibbo-Kobayashi-Maskawa mixing is
neglected. Then, the fermion masses are
\begin{equation}
m_f =\frac{2m_W}{g}\left(G_{f,1}\cos\beta
+G_{f,2}\sin\beta\right).
\end{equation}

As was shown by Glashow and Weinberg \cite{FCNC}, the tree-level
FCNC's mediated by Higgs bosons are absent in case when all
fermions of a given electric charge couple to no more than one
Higgs doublet. This restriction leads to four different models, as
discussed in Ref. \cite{santos}. In what follows, we will use the
most general parametrization (\ref{L:Yuk}) including the models
mentioned as well as other possible variations of the Yukawa
sector without the tree-level FCNC's.

The gauge sector is taken to be
\begin{eqnarray}
\mathcal{L}_{\mathrm{gauge}} & = & - \frac{1}{4}{F}^{\mu\nu}{F}_{\mu\nu} - \frac{1}{4}{F}^{\mu\nu}_a {F}_{\mu\nu}^{a} - \frac{1}{4}{\tilde{F}}^{\mu\nu}{\tilde{F}}_{\mu\nu}.
\end{eqnarray}

In the present paper all calculations are carried out in the
Feynman-'t Hooft gauge, the gauge-fixing functions are
\begin{eqnarray}
\label{L:gauge_fixing_functions}
G^a & = & \frac{1}{\sqrt{\xi}} \left( \partial^\mu A_\mu^a + \xi \frac{ig}{2}  \sum_{i=1}^{2} \left( \varphi^{\dagger}_i \sigma^a \varphi_{0i} - \varphi_{0i}^\dagger \sigma^a \varphi_i \right) \right) , \nonumber\\
G & = & \frac{1}{\sqrt{\xi}} \left( \partial^\mu B_\mu + \xi \frac{ig'}{2}  \sum_{i=1}^{2} \left( \varphi^{\dagger}_i  \varphi_{0i} - \varphi_{0i}^\dagger \varphi_i \right) \right) , \nonumber\\
\tilde{G} & = & \frac{1}{\sqrt{\xi}} \left( \partial^\mu \tilde{B}_\mu + \xi \frac{i\tilde{g}}{2}  \sum_{i=1}^{2} \left( \varphi^{\dagger}_i  \varphi_{0i} - \varphi_{0i}^\dagger \varphi_i \right) \right) , \nonumber\\
\varphi_{0i} & = &\left( \begin{array}{c}0 \\ v_{i}/\sqrt{2} \end{array} \right) .
\end{eqnarray}
Then, the gauge-fixing part of the Lagrangian reads
\begin{eqnarray}
\mathcal L_{\mathrm{gauge\,fixing}} & = & - \frac{1}{2}\left(\sum_{a=1}^3 G^{a 2} + G^2 + \tilde{G}^2 \right).
\end{eqnarray}
The kinetic part of Faddeev-Popov sector is
\begin{eqnarray}
\mathcal L_{\mathrm{ghost}} & = & - \bar{u}^+ (\partial^2 + \xi m_{W}^2) u^- - \bar{u}^- (\partial^2 + \xi m_{W}^2) u^+ \nonumber\\
&& - \bar{u}_Z (\partial^2 + \xi m_{Z}^2) u_Z - \bar{u}_A \partial^2 u_A \nonumber\\
&& - \bar{u}_{Z'} (\partial^2 + \xi m_{Z'}^2) u_{Z'},
\end{eqnarray}
$\xi$ is  gauge-fixing parameter. For arbitrary $\xi$ the gauge-boson propagator is
\begin{eqnarray}
iD^{\mu\nu}(p) = - \frac{i}{p^2-m^2+i\epsilon}  \left( g^{\mu\nu}+(\xi-1)\frac{p^\mu p^\nu}{p^2-\xi m^2} \right) .
\end{eqnarray}

The MSM parametrization can be obtained by putting $\tan\beta$,
$\tan \alpha$, $\mu_2$, $\lambda_{2,3,4,5}$, $\tilde{Y}_{\phi_2,1,2}$,
$G_{d,2}$ and $G_{u,2}$ to zero and dropping the
summations over $i$ in (\ref{L:scalar}), (\ref{L:Yuk}) and
(\ref{L:gauge_fixing_functions}).

In Refs. \cite{Gulov:2000eh,Gulov:1999ry} the relations between
$Z'$ parameters were found from the requirement that the
underlying extended model is a renormalizable one. They read
\begin{equation} \label{rgr2}
\tilde{Y}_{\phi,1} = \tilde{Y}_{\phi,2} \equiv \tilde{Y}_{\phi},
\qquad \tilde{Y}_{L,f}= \tilde{Y}_{L,f^*},\qquad \tilde{Y}_{R,f} =
\tilde{Y}_{L,f} + 2 T_{3f}~ \tilde{Y}_{\phi}
\end{equation}
in case of Abelian $Z'$ boson. Here $f$ and $f^*$ are the partners
of the $SU(2)_L$ fermion doublet ($l^* = \nu_l, \nu^* = l, q^*_u =
q_d$ and $q^*_d = q_u$), $T_{3f}$ is the third component of weak
isospin. These relations are used all over the present paper. They
can be also rewritten in terms of vector couplings, axial-vector
couplings, and the $Z$--$Z'$ mixing angle.

%\section{Bibliography}

\end{document}